\documentclass[aps,prl,nobibnotes,nofootinbib,showpacs,reprint]{revtex4-1}

\usepackage{graphicx}
\usepackage{amsmath} 	

\newcommand{\vect}[1]{\mathbf{#1}}

\newcommand{\Fig}[1]{FIG.~\ref{#1}}
 
\newcommand{\Eq}[1]{Eq.~(\ref{#1})} 
\newcommand{\Eqs}[2]{Eqs.~(\ref{#1})~and~(\ref{#2})} 
\newcommand{\Ep}{\tilde{\mathcal{E}_{p}}}
\newcommand{\Ee}{\tilde{\mathcal{E}_{e}}}
\newcommand{\Eg}{\tilde{\mathcal{E}_{g}}}
\newcommand{\we}{\omega_{e}}
\newcommand{\wg}{\omega_{g}}
\newcommand{\chie}{\chi_{e}}
\newcommand{\chig}{\chi_{g}}
\newcommand{\Dw}{\Delta \omega}
\newcommand{\Sw}{\Sigma \omega}
\newcommand{\Dv}{\Delta}
\newcommand{\Dvs}{\Delta_{s}}

\newcommand{\Wm}{\Omega}
\newcommand{\Wms}{\Omega_{s}}

\newcommand{\aue}{\hat{a}_{e}^{\dagger}}
\newcommand{\ade}{\hat{a}_{e}}
\newcommand{\aug}{\hat{a}_{g}^{\dagger}}
\newcommand{\adg}{\hat{a}_{g}}

\newcommand{\alp}{\alpha}
\newcommand{\bta}{\beta}
\newcommand{\alps}{\alpha_{s}}
\newcommand{\btas}{\beta_{s}}

\begin{document}

\title{Dualism between Optical and Difference Parametric Amplification}

\author{Wayne~Cheng-Wei~Huang}
\affiliation{The Institute for Quantum Science and Engineering, Texas A\&M University, College Station, Texas 77843, USA }
\affiliation{Department of Physics and Astronomy, University of Nebraska-Lincoln, Lincoln, Nebraska 68588, USA }

\author{Herman~Batelaan}
\affiliation{Department of Physics and Astronomy, University of Nebraska-Lincoln, Lincoln, Nebraska 68588, USA }
\email{email: hbatelaan2@unl.edu}

\begin{abstract}

Breaking the symmetry in a coupled wave system can result in unusual amplification behavior. In the case of difference parametric amplification the resonant pump frequency is equal to the difference, instead of the sum, frequency of the normal modes. We show that sign reversal in the symmetry relation of parametric coupling give rise to difference parametric amplification as a dual of optical parametric amplification. For optical systems, our result can potentially be used for efficient XUV amplification.

\end{abstract}


\maketitle 

Parametric processes are essential to quantum optical applications including frequency conversion, quantum communication, and nonclassical state generation \cite{Dutt, Ma, Meekhof, Vlastakis}. In particular, the application of squeezed light in precision measurement has led to enhanced sensitivity for gravitational wave detection \cite{LIGO}. Parametric interaction occurs when driving a nonlinear dipole with two frequency inputs. In a doubly resonant cavity, two non-degenerate target frequencies, $\we > \wg$, can be parametrically coupled to a pump frequency through a nonlinear medium \cite{Colville, Rivoire}. When the pump frequency $\nu$ is equal to the sum-frequency $\Sw \equiv \we + \wg$ or the difference-frequency $\Dw \equiv \we - \wg$, resonant parametric interaction occurs. A sum-frequency will facilitate energy transfer from the pump field $E_{p}$ to the target fields $E_{e,g}$, leading to amplification. A difference-frequency will promote energy exchange between the target fields without changing their total energy \cite{Boyd}. In the framework of quantum optics, the former corresponds to anti-Jaynes-Cummings interaction and the latter amounts to Jaynes-Cummings interaction \cite{Kippenberg, Gerry}. 
 
In a recent proposal by Svidzinsky et \textit{al.} \cite{Svidzinsky1}, a semiclassical approach was used to show that Jaynes-Cummings interaction could lead to strong amplification of light in a superradiant atomic gas, if such a coupled system is driven with an external difference-frequency pump. This quickly leads to the conceptual difficulty that energy conservation is violated. In optical parametric amplification (OPA) energy transfers from the pump field to the target fields because one sum-frequency photon, having higher energy, breaks into two target-frequency photons with smaller energy \cite{Gerry}. In the case where the difference-frequency pump drives the amplification, such a photon picture cannot apply since the energy of one difference-frequency photon is less than the total energy of two target-frequency photons. Assuming that this effect exists, what is then the mechanism for energy transfer? To shed light on this puzzle, we turn to Maxwell equations where OPA was originally studied \cite{Kroll, Tien, Giordmaine}. 

In this Letter, we show that difference parametric amplification (DPA), i.e. amplification based on a difference-frequency drive, does not violate energy conservation at the level of classical physics. We illustrate the dualism between DPA and OPA through the symmetry relation of parametric coupling. Given that quantum mechanics is a more superior theory than classical mechanics, a corresponding quantum mechanism should exist. We argue that the combination of DPA and the Jaynes-Cummings Hamiltonian will lead to non-Hermiticity. This gives rise to complex-valued expectation values and may explain why the photon picture does not apply for DPA. 

We note that DPA, if realized, presents potential advantages for delivering efficient XUV amplification. The state-of-the-art upconversion light sources are based on either multiphoton excitation or higher-harmonic generation \cite{Zhou, Haase, Ghimire}. These processes suffer from deteriorating conversion efficiency as the target frequency gets into the ultraviolet regime \cite{Wang, Ferray, Lein}. In contrast, DPA remains as a first-order nonlinear process regardless of how high the target frequency is. This feature renders DPA a potential mechanism for efficient amplification in the XUV regime.

\begin{figure}[b]
\centering
\scalebox{0.4}{\includegraphics{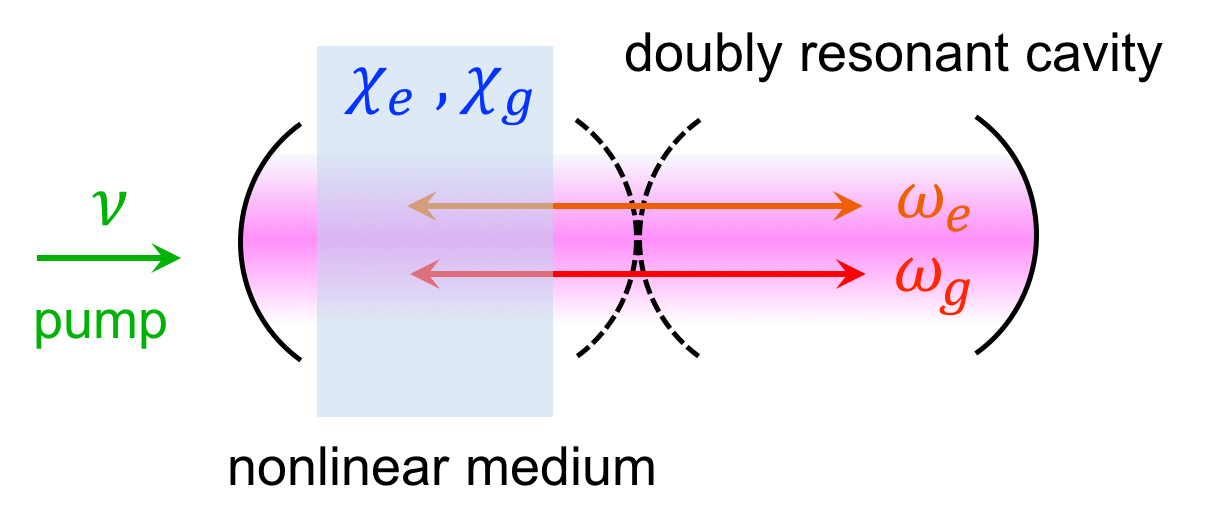}}
\caption{Parametric pumping for two coupled cavities. The transmissivity of the coupling mirrors (dashed line) determines the strength of normal mode splitting, hence the difference-frequency between the two cavity modes $\Dw \equiv \we-\wg$. The nonlinear medium (blue) is assumed to mediate parametric interaction between the cavity modes $E_{e,g}$ and the pump $E_{p}$ with nonlinear coupling parameters $\chi_{e,g}$.}
\label{fig:OPO}
\end{figure}

To illustrate the concept of DPA and its connection to OPA, we start with Maxwell equations for waves in a nonlinear medium \cite{Boyd},
\begin{equation}\label{maxwell}
	\nabla^2\vect{E} - \frac{\epsilon^{(1)}}{c^2}\frac{\partial^2\vect{E}}{\partial t^2} = \frac{4\pi}{c^2}\frac{\partial^2 \vect{P}^{\textrm{NL}}}{\partial t^2}
\end{equation}
where $\epsilon^{(1)}$ represents the linear dielectric response of the medium which, for simplicity, is assumed to be isotropic dispersionless. The dipole moment of the nonlinear medium $\vect{P}^{\textrm{NL}}$ acts as a driving source and couples the target field $\vect{E}$ with a pump field $\vect{E}_{p}$ through $\vect{P}^{\textrm{NL}} = \chi^{(2)}\vect{E}_{p}\vect{E}$, where $\chi^{(2)}$ is a dielectric tensor that characterizes the second-order nonlinear response of the medium. Based on \Eq{maxwell}, we consider the wave dynamics of two eigenmodes $E_{e,g}$ of frequencies $\omega_{e,g}$ in a doubly resonant cavity (See \Fig{fig:OPO}). We assume small normal mode splitting compared to the eigenfrequencies, $0 < \we-\wg \ll \omega_{e,g}$. Through a second-order nonlinear medium, the two target fields $E_{e,g}$ are parametrically coupled by an injected pump field $E_{p}$. The coupled wave equations can be derived from \Eq{maxwell},
\begin{equation}\label{main_eq1}
	\begin{split}
	\left\{
	\begin{array}{l}
	\displaystyle \frac{d^2 E_{e}}{dt^2} = -\we^2 E_{e}+ \chig E_{g}E_{p}	\\	\\
	\displaystyle \frac{d^2 E_{g}}{dt^2} = -\wg^2 E_{g} + \chie E_{e}E_{p}			
	\end{array}
	\right. 	,
	\end{split}
\end{equation}
where $\chig$ and $\chie$ are the nonlinear coupling parameters for $E_{g}$ and $E_{e}$ respectively. Conventionally, the nonlinear coupling is symmetric with respect to the target frequencies, $\chig = \chie$. However, here we make the distinction and extend the analysis to the more general case where the two coupling parameters can be made different, $\chig \ne \chie$. In addition, we remark that \Eq{main_eq1} is the diagonalized representation for all parametrically coupled systems, including the cases discussed in reference \cite{Svidzinsky1, Svidzinsky2, Chen, Zhang}. Given the pump field $E_{p}(t) = A_{0}\cos{(\nu t + \phi)}$, the coupled equations can be transformed with the complex notation $E = (\tilde{E} +\tilde{E}^{\ast})/2$ and rotating-wave frame $\tilde{E}(t) = \tilde{\mathcal{E}}(t) e^{-i\omega t}$ to 
\begin{widetext}
\begin{equation}\label{main_eq1-2}
	\begin{split}
	\left\{
	\begin{array}{l}
	\displaystyle  \frac{d^2}{dt^2}\left( \Ee e^{-i\we t} +  \Ee^{\ast} e^{i\we t} \right) + \we^2 \left( \Ee e^{-i\we t} +  \Ee^{\ast} e^{i\we t} \right) = \frac{\chig}{2} \left( \Eg e^{-i\wg t} +  \Eg^{\ast} e^{i\wg t}  \right)\left( \Ep e^{-i\nu t} +  \Eg^{\ast} e^{i\nu t}  \right)  	\\	\\
	\displaystyle  \frac{d^2}{dt^2}\left( \Eg e^{-i\wg t} +  \Eg^{\ast} e^{i\wg t} \right) + \wg^2 \left( \Eg e^{-i\wg t} +  \Eg^{\ast} e^{i\wg t} \right) = \frac{\chie}{2} \left( \Ee e^{-i\we t} +  \Ee^{\ast} e^{i\we t}  \right)\left( \Ep e^{-i\nu t} +  \Eg^{\ast} e^{i\nu t}  \right)  
	\end{array}
	\right. 	,
	\end{split}
\end{equation}
\end{widetext}
where $\Ep \equiv A_{0} e^{-i\phi}$ is the pump amplitude. To simplify the above equations, we eliminate the non-resonant terms with the rotating-wave approximation. Also, we will use the slow-varying approximation, $d\mathcal{E}_{e,g}(t)/dt \ll \omega_{e,g}\mathcal{E}_{e,g}(t)$, to reflect the slow-varying envelop $\mathcal{E}_{e,g}(t)$ and focus only on the fast dynamics at the timescales $1/\omega_{e,g}$. Under these assumptions, \Eq{main_eq1-2} becomes
\begin{equation}\label{main_eq2}
	\begin{split}
	\left\{
	\begin{array}{l}
	\displaystyle  \frac{d\Ee}{dt} = \frac{i\chig}{4\we} \left( \Eg\Ep e^{-i\Dv t} +  \Eg^{\ast}\Ep e^{-i\Dvs t} \right)  	\\	\\
	\displaystyle  \frac{d\Eg}{dt} = \frac{i\chie}{4\wg} \left( \Ee\Ep^{\ast} e^{i\Dv t} +  \Ee^{\ast}\Ep e^{-i\Dvs t} \right)
	\end{array}
	\right. 	,
	\end{split}
\end{equation}
where $\Dv \equiv \nu - \Dw$ and $\Dvs \equiv \nu - \Sw$ are the pump detunings from the difference-frequency $\Dw \equiv \we - \wg$ and the sum-frequency $\Sw \equiv \we + \wg$, respectively. Later, the validity of the approximations will be shown by the agreement between the analytical solution and the numerical simulation of \Eq{main_eq1}. 

In OPA, the pump frequency is close to the sum-frequency $\nu \approx \Sw$ $(|\Dv| \gg 0)$, and \Eq{main_eq2} can be further simplified by making again the rotating-wave approximation,
\begin{equation}\label{OPA_eq}
	\left\{
	\begin{array}{l}
	\displaystyle  \frac{d\Ee}{dt} = \alps \Eg^{\ast} e^{-i\Dvs t} 	\\	\\
	\displaystyle  \frac{d\Eg}{dt} = \btas \Ee^{\ast} e^{-i\Dvs t} 
	\end{array}
	\right. ,
\end{equation}
where the gain parameters are defined as $\alps \equiv i\chig\Ep/4\we$ and $\btas \equiv i\chie\Ep/4\wg$. The target field solutions $\tilde{E}_{e,g}(t)$ can be derived accordingly, 
\begin{equation}\label{OPA_sol}
	\begin{split}
	 \tilde{E}_{e}(t)  =& \left\{ \tilde{E}_{e}(0) \left[ \cosh{\left( \frac{\Wms t}{2} \right)} + \frac{i\Delta_{s}}{\Wms}\sinh{\left( \frac{\Wms t}{2} \right)} \right] \right.	\\
	 		&\left. + \tilde{E}_{g}^{\ast}(0) \frac{2\alp_{s}}{\Wms}\sinh{\left( \frac{\Wms t}{2} \right)} \right\}e^{-i(\we+\Delta_{s}/2)t}	,\\	
	\tilde{E}_{g}(t)  =& \left\{ \tilde{E}_{g}(0) \left[ \cosh{\left( \frac{\Wms^{\ast} t}{2} \right)} + \frac{i\Delta_{s}}{\Wms^{\ast}}\sinh{\left( \frac{\Wms^{\ast} t}{2} \right)} \right] \right. 	\\
 		&\left.  + \tilde{E}_{e}^{\ast}(0) \frac{2\bta_{s}}{\Wms^{\ast}}\sinh{\left( \frac{\Wms^{\ast} t}{2} \right)} \right\}e^{-i(\wg+\Delta_{s}/2)t}		, 
	\end{split}
\end{equation}
where the OPA gain rate is $\Wms = \sqrt{-\Dvs^2 + 4\alps\btas^{\ast}}$. The analytical solution to \Eq{main_eq1} is thus $E_{e,g}(t)= (\tilde{E}_{e,g}(t) +\tilde{E}^{\ast}_{e,g}(t))/2$. Seeing from \Eq{OPA_sol}, we notice that the dynamic behavior of the coupled wave system is fully determined by what we call the symmetry relation herein, the sign of  $\chie\chig$. Assuming a sufficiently strong pump, $|\Ep| > 4|\Dvs|\sqrt{\we\wg/|\chie\chig|}$, the positive symmetry relation $\chie\chig > 0$ implies that $\alps\btas^{\ast} = \chie\chig|\Ep|^2/16\we\wg > 0$. This guarantees a real-valued OPA gain rate, $\Wms \in \Re$, and gives rise to exponential amplification of the target fields under a sum-frequency pump, as expected for OPA.

\begin{figure}[t]
\centering
\scalebox{0.37}{\includegraphics{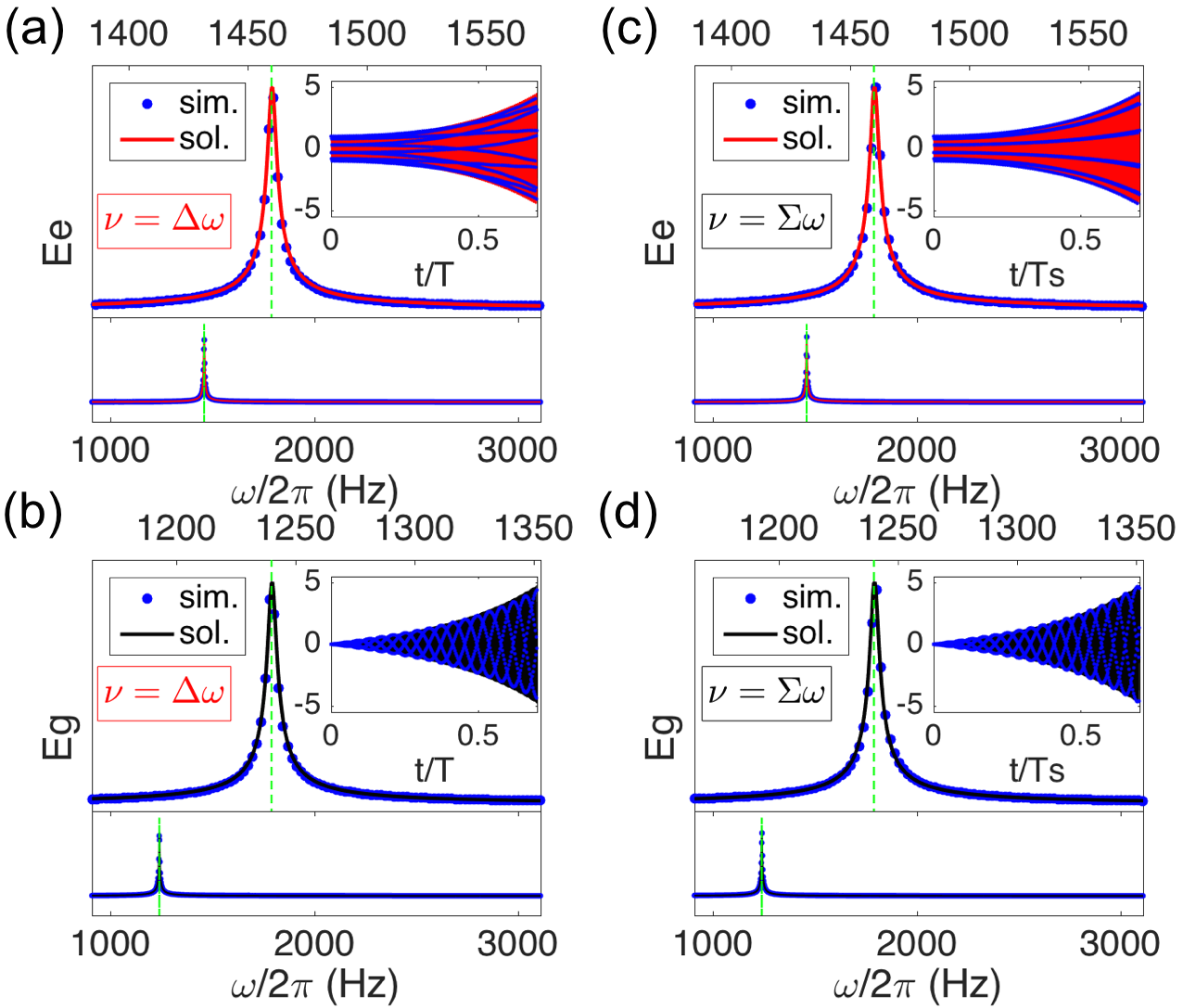}}
\caption{Comparison between analytical solutions and simulations for amplification via parametric pumping. (a)(b) Provided a negative symmetry relation $\chie\chig < 0$, amplification of the target fields $E_{e,g}$ can only be achieved through a difference-frequency pump $\nu = \Dw$. The upper right insets give the temporal evolution of the target fields. Here, initial conditions $\tilde{E}_{e}(0) = 1$ and $\tilde{E}_{g}(0) = 0$ are assumed. Fourier spectra of the fields show a single spectral peak at the respective frequencies $\we/2\pi = 1460$ Hz and $\wg/2\pi = 1240$ Hz (bottom panels). The width of the spectral peaks characterizes the exponential growth rate of the field amplitude (top panels). A good agreement is found between the analytical solutions (black and red) and the simulation (blue). (c)(d) Provided a positive symmetry relation $\chie\chig > 0$, amplification can only be attained through a sum-frequency pump $\nu = \Sw$. The temporal behavior and spectral property are similar to the case of a difference-frequency pump. The temporal evolution is plotted at the timescale $T \equiv 2\pi/\Wm$ and $T_{s} \equiv 2\pi/\Wms$ for DPA and OPA respectively.}
\label{fig:FFT_traj}
\end{figure}

\begin{figure}[t]
\centering
\scalebox{0.37}{\includegraphics{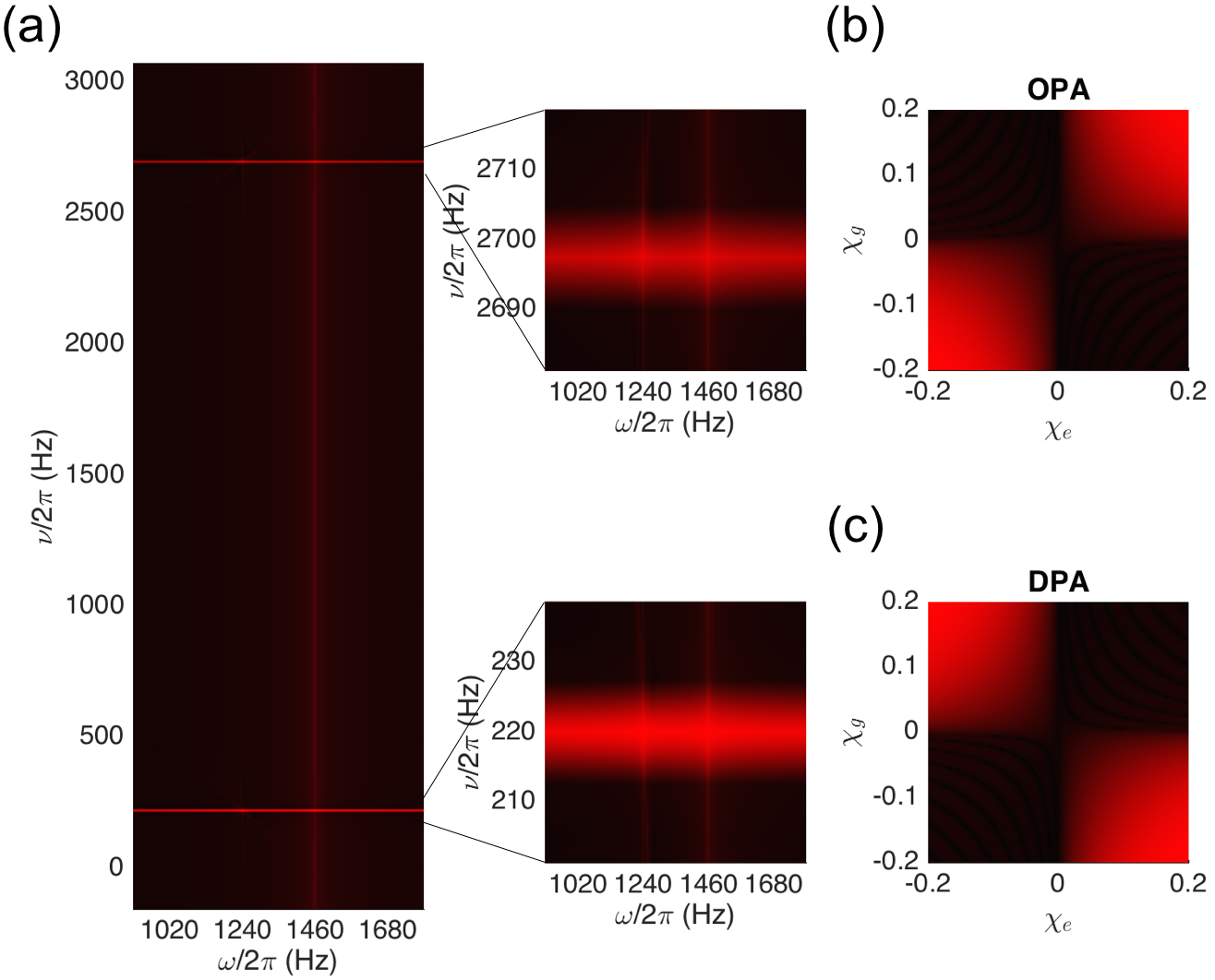}}
\caption{Parametric resonance and Fourier spectra for OPA and DPA. (Field amplitude in logarithmic color scale) (a) Pronounced amplification of the two target frequencies $\we/2\pi = 1460$ Hz and $\wg/2\pi = 1240$ Hz appear for resonant pumping at the sum-frequency $\nu/2\pi = 2680$ Hz (OPA) and the difference-frequency pump at $\nu/2\pi = 220$ Hz (DPA). (b)(c) The parameter regimes $(\chie, \chig)$ for OPA and DPA are mutually exclusive. Initial conditions $\tilde{E}_{e}(0) = 1$ and $\tilde{E}_{g}(0) = 0$ are assumed}
\label{fig:drivingw_FFT}
\end{figure}

When a difference-frequency pump $\nu \approx \Dw$ is used instead, the coupled equations in \Eq{main_eq2} become
\begin{equation}\label{DPA_eq}
	\left\{
	\begin{array}{l}
	\displaystyle  \frac{d\Ee}{dt} = \alp \Eg e^{-i\Dv t} 	\\	\\
	\displaystyle  \frac{d\Eg}{dt} = \bta \Ee e^{i\Dv t} 
	\end{array}
	\right. .
\end{equation}
The use of a difference-frequency pump results in a new set of gain parameters $\alp \equiv i\chig\Ep/4\we$ and $\bta \equiv i\chie\Ep^{\ast}/4\wg$, and the solutions for the target fields $\tilde{E}_{e,g}(t)$ are
\begin{equation}\label{DPA_sol}
	\begin{split}
	 \tilde{E}_{e}(t)  =& \left\{ \tilde{E}_{e}(0) \left[ \cosh{\left( \frac{\Wm t}{2} \right)} + \frac{i\Dv}{\Wm}\sinh{\left( \frac{\Wm t}{2} \right)} \right] \right.	\\
	 		&\left. + \tilde{E}_{g}(0) \frac{2\alp}{\Wm}\sinh{\left( \frac{\Wm t}{2} \right)} \right\}e^{-i(\we+\Dv/2)t}	,\\	
	\tilde{E}_{g}(t)  =& \left\{ \tilde{E}_{g}(0) \left[ \cosh{\left( \frac{\Wm t}{2} \right)} - \frac{i\Dv}{\Wm}\sinh{\left( \frac{\Wm t}{2} \right)} \right] \right. 	\\
 		&\left.  + \tilde{E}_{e}(0) \frac{2\bta}{\Wm}\sinh{\left( \frac{\Wm t}{2} \right)} \right\}e^{-i(\wg-\Dv/2)t}	,
	\end{split}
\end{equation}
where the DPA gain rate is $\Wm = \sqrt{-\Dv^2+4\alp\bta}$. The important difference in the gain parameters $\alp$ and $\bta$ makes it possible to attain amplification through a difference-frequency pump and a negative symmetry relation $\chie\chig < 0$ ($\alp\bta = -\chie\chig|\Ep|^2/16\we\wg > 0$). With a sufficiently strong pump, $|\Ep| > 4|\Dv|\sqrt{\we\wg/|\chie\chig|}$, \Eq{DPA_sol} implies that the target field can be exponentially amplified with a real-valued DPA gain rate, $\Wm \in \Re$.

We compare the analytical solutions, \Eqs{OPA_sol}{DPA_sol}, to the simulation results of \Eq{main_eq1}, assuming the positive symmetry relation for $\nu = \Sw$ and the negative symmetry relation for $\nu = \Dw$. The good agreement in both cases justifies the use of rotating-wave approximation and the slow-varying approximation in the analysis (See \Fig{fig:FFT_traj}). Without loss of generality, the target frequencies are taken to be $\wg/2\pi = 1240$ Hz and $\we/2\pi = 1460$ Hz from acoustic waves. This makes the simulation less stiff as the ratio between the target frequency and the difference-frequency $\Dw/2\pi = 220$ Hz is kept within 10. Generally, the solutions can be applied to any frequency regime.

\begin{figure}[t]
\centering
\scalebox{0.37}{\includegraphics{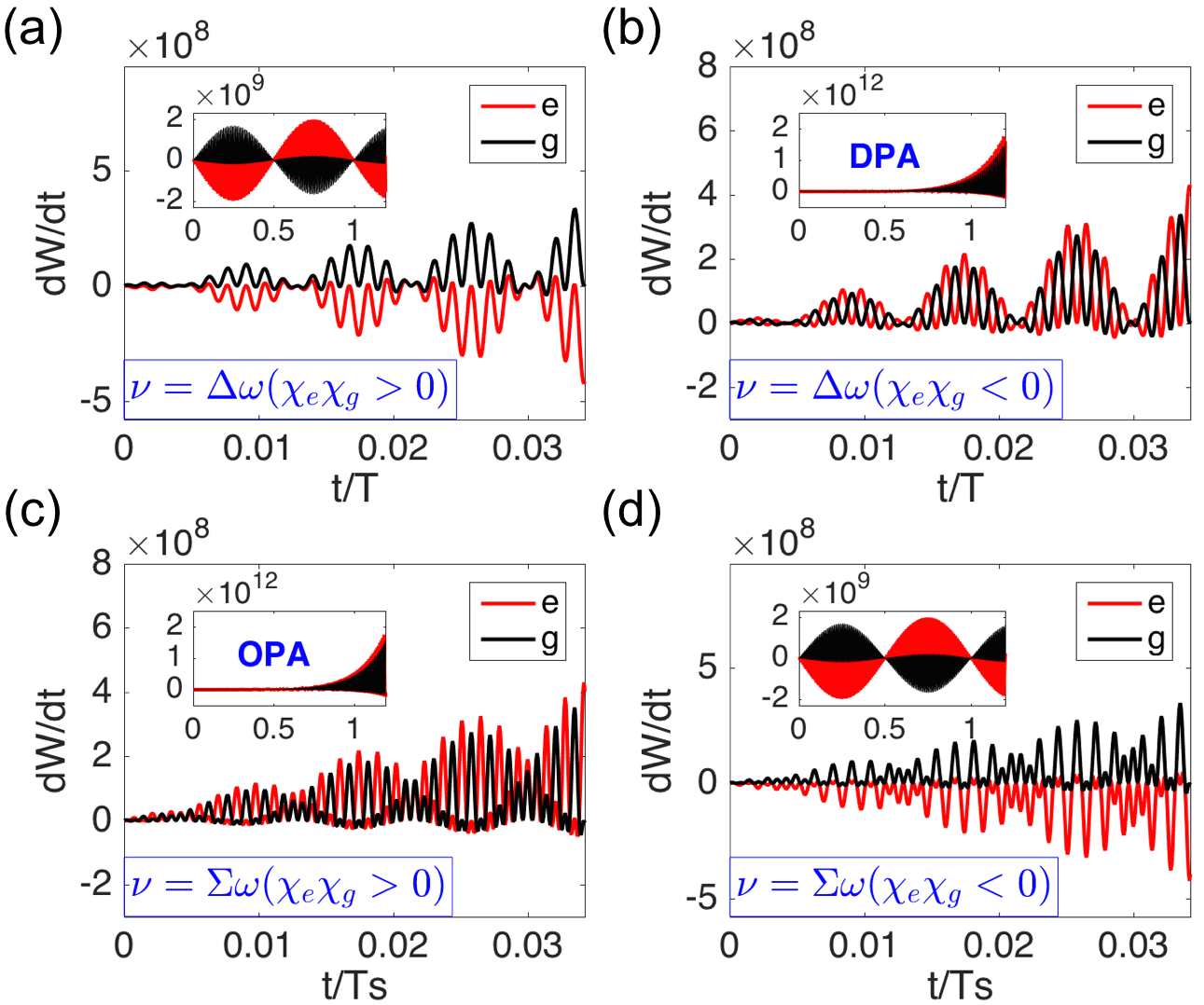}}
\caption{Temporal evolution of energy flow in OPA and DPA. (a) Provided a positive symmetry relation $\chie\chig > 0$, a difference-frequency pump $\nu = \Dw$ promotes energy exchange between the target fields $E_{e,g}$. Energy flows in the two fields have opposite signs. A positive sign represents energy gain; a negative sign represents energy loss. The negative energy flow is flipped to positive as the energy of the respective field is depleted (inset). Initial conditions $\tilde{E}_{e}(0) = 1$ and $\tilde{E}_{g}(0) = 0$ are assumed. (b) Under a negative symmetry relation $\chie\chig < 0$, a difference-frequency pump $\nu = \Dw$ can cause amplification for the two target fields (DPA). The total energy of the fields increases over time. (c) With the same nonlinear coupling parameters as (a), a sum-frequency pump $\nu = \Sw$ can cause field amplification as in the case of (b) (OPA). (d) Given the same nonlinear coupling parameters as (c), a sum-frequency pump $\nu = \Sw$ will induce energy exchange between the two target fields as in the case of (a). These four scenarios indicate that the roles of a difference-frequency pump and a sum-frequency pump are exchanged in the two mutually exclusive parameter regimes $\chie\chig > 0$ and $\chie\chig < 0$.}
\label{fig:work_traj}
\end{figure}

While the resonant frequencies for OPA ($\nu = \Sw$) and DPA ($\nu = \Dw$) are vastly apart, both give rise to amplification of the same target frequencies $\omega_{e,g}$ with mutually exclusive parameter regimes, $\chie\chig > 0$ and $\chie\chig < 0$ (See \Fig{fig:drivingw_FFT}). The dualism between OPA and DPA is made clear when considering the energy flows in the coupled wave system. Using \Eq{main_eq1}, the energy transfer to a field can be calculated through the driving term $\chi_{e,g}E_{e,g}E_{p}$,
\begin{equation}\label{DPA_work0}
	W_{e,g} = \epsilon_{0}E^2_{e,g} = \epsilon_{0}\int dt \left( \frac{dE_{e,g}}{dt}\chi_{e,g}E_{e,g}E_{p} \right)	. 
\end{equation} 
According to \Eq{DPA_sol}, a difference-frequency pump $\nu = \Dw$ with positive symmetry relation $\chie\chig > 0$ gives the solution
\begin{equation}\label{DPA_work1}
	\begin{split}
	E_{e}(t) =& Q_{e}(0)\cos{(\we t)}\cos{(bt/2)},		\\
	E_{g}(t) =& \frac{\chie A_{0}Q_{e}(0)}{2\wg b}\sin{(\wg t)}\sin{(bt/2)},
	\end{split}	
\end{equation}
where $\phi = 0$ is assumed. The initial conditions are set to be $\tilde{E}_{e}(0) = Q_{e}(0)$ and $\tilde{E}_{g}(0) = 0$. Parameters $a$ and $b$ are defined as the real and imaginary parts of the DPA gain rate $\Wm = a + i b$. With the positive symmetry relation $\chie\chig > 0$, it follows that $a = 0$ but $b \neq 0$. The energy flow in the fields can be subsequently  computed with the slow-varying approximation,
\begin{equation}\label{DPA_work3}
	\begin{split}
	\frac{dW_{e}(t)}{dt} \approx& -\epsilon_{0}\chie\chig \left( \frac{ A_{0}^2 Q_{e}^2(0)}{4b} \right) \left( \frac{\we}{\wg} \right)\sin{\left( \we t \right)}\sin{\left( \wg t \right)}	\\
					& \times \cos{\left(\Dw t \right)} \sin{\left( b t\right)},	\\
	\frac{dW_{g}(t)}{dt} \approx& \epsilon_{0}\chie^2\left( \frac{ A_{0}^2 Q_{e}^2(0)}{4b} \right) \cos{\left( \we t \right)}\cos{\left( \wg t \right)}	\\
					& \times \cos{\left(\Dw t \right)}\sin{\left( b t\right)}	.
	\end{split}
\end{equation}
As shown in \Fig{fig:work_traj}(a), the energy flow in the two target fields have similar strength but opposite sign, implying that energy is exchanged between the two fields.  When the energy of a field is depleted, the sign of its energy flow is reversed. The depletion rate is characterized by the imaginary part of the DPA gain rate $b$. 

In the case of DPA (negative symmetry relation $\chie\chig < 0$), the field solutions are
\begin{equation}
	\begin{split}
	E_{e}(t) =& Q_{e}(0)\cos{(\we t)}\cosh{(at/2)},		\\
	E_{g}(t) =& \frac{\chie A_{0}Q_{e}(0)}{2\wg a}\sin{(\wg t)}\sinh{(at/2)}.
	\end{split}
\end{equation}
The sinusoidal functions in  \Eq{DPA_work1} are now replaced by hyperbolic functions because $a \neq 0$ but $b = 0$. The corresponding energy flows are 
\begin{equation}
	\begin{split}\label{DPA_work2}
	\frac{dW_{e}(t)}{dt} \approx& -\epsilon_{0}\chie\chig \left( \frac{A_{0}^2 Q_{e}^2(0)}{4a} \right) \left( \frac{\we}{\wg} \right) \sin{\left( \we t \right)}\sin{\left( \wg t \right)}	\\
					& \times \cos{\left(\Dw t \right)}\sinh{\left( a t\right)},	\\
	\frac{dW_{g}(t)}{dt} \approx& \epsilon_{0}\chie^2 \left( \frac{A_{0}^2 Q_{e}^2(0)}{4a} \right) \cos{\left( \we t \right)}\cos{\left( \wg t \right)}	\\
					& \times \cos{\left(\Dw t \right)}\sinh{\left( a t\right)}	.
	\end{split}
\end{equation}
The negative symmetry relation makes the energy flows in both target fields obtain a positive sign, leading to simultaneous excitation of the two fields (See \Fig{fig:work_traj}(b)). The modulation term $\cos{\left( \Dw t \right)}$ indicates that energy is pumped into the target fields at the rate of the difference-frequency. The fast oscillations in the energy flow are out of phase, suggesting that the fields take turns to draw energy from the pump. The hyperbolic term $\sinh{\left( a t\right)}$ shows the exponential energy growth in the two fields at the rate $a$, which is the real part of the DPA gain rate. 

Remarkably, the dynamic behavior of the target fields in the two parameter regimes ($\chie\chig > 0$ and $\chie\chig <0$) are reversed if the coupled wave system is provided with a sum-frequency pump $\nu = \Sw$. Under the positive symmetry relation $\chie\chig > 0$, as in the scheme of OPA, the system will undergo the same amplification as described by \Eq{DPA_work2} with the parameter $a$ replaced by the real part of the OPA gain rate, $a_{s} = Re(\Wms)$ (See \Fig{fig:work_traj}(c)). When the symmetry relation is negative $\chie\chig < 0$, energy is exchanged between the two target fields in a conservative way, as shown in \Fig{fig:work_traj}(d). This behavior can be described by \Eq{DPA_work3} with the parameter $b$ replace by the imaginary part of the OPA gain rate, $b_{s} = Im(\Wms)$.

\begin{figure*}[t]
\centering
\scalebox{0.38}{\includegraphics{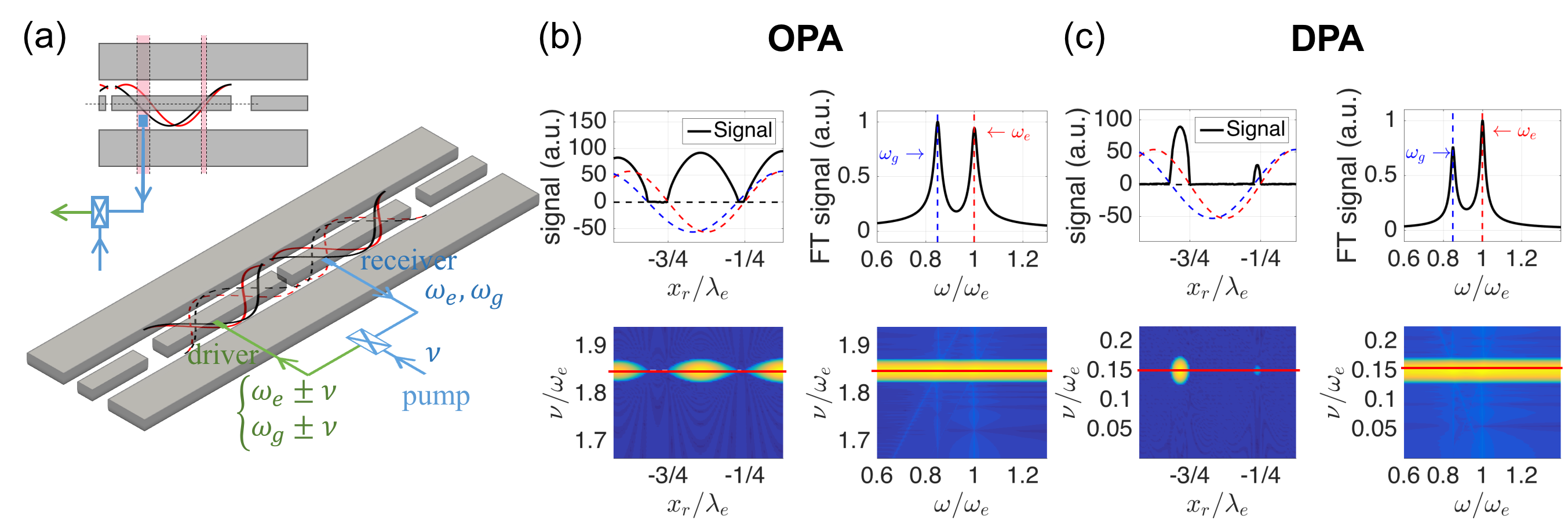}}
\caption{Theoretical demonstration of transition from OPA to DPA with a change in a single physical parameter $x_{r}$. (a) A coupled coplanar waveguide cavity is doubly-resonant at $\omega_{e,g}$. Parametric pumping is achieved through a feedback loop in two steps. First, cavity field signals $\omega_{e,g}$ are passed from a receiver antenna to an ideal mixer  to be mixed with a pump signal $\nu$. Second, the output signal $\omega_{e,g} \pm \nu$ of the mixer is sent to a driver antenna to pump the coupled cavity. The inset shows the spatial phase profiles of the cavity modes $\cos{(k_{e}x_{r})}$ (red) and $\cos{(k_{g}x_{r})}$ (black). When the receiver antenna (blue) is parked in the regimes in-between the nodes (pink), parametric coupling via the feedback loop will take a negative symmetry relation $\chie\chig < 0$, which facilitates DPA. (b) Pumping the cavity at the sum-frequency $\nu = \Sw$ shows amplification in the regimes of positive symmetry $\chie\chig > 0$. (c) Pumping the cavity at the difference-frequency $\nu = \Dw$ gives rise to field amplification in the regimes in-between the nodes, as indicated in the inset of (a). Initial conditions $\tilde{E}_{e}(0) = 1$ and $\tilde{E}_{g}(0) = 0$ are assumed.}
\label{fig:cavity}
\end{figure*}

The four scenarios summarized in \Fig{fig:work_traj} illustrate the dualism between OPA and DPA. The sign reversal in the symmetry relation switches the roles of a sum-frequency pump and a difference-frequency pump in the coupled wave system. While the positive symmetry relation $\chie\chig > 0$ promotes OPA, the negative symmetry relation $\chie\chig < 0$ facilitates DPA. The symmetry relation reflects the symmetry built into the coupling mechanism. To provide a physical context for discussion, we devise a thought experiment where transition between OPA and DPA is controlled by a single knob. In \Fig{fig:cavity}(a), two identical microwave coplanar waveguide cavities are capacitively coupled. Normal mode splitting makes two close eigenfrequencies $\omega_{e,g}$. Parametric pumping for the cavity modes is provided by a feedback loop in two steps. First, the field signal $E_{e}(x_{r},t) + E_{g}(x_{r},t)$ is taken from a receiver antenna sitting at $x_{r}$ and mixed with a pump signal $E_{p} = A_{0}\cos{(\nu t + \phi)}$ through an ideal mixer of output efficiency $\chi$. Second, the output signal from the mixer is fed to a driver antenna as the pump for the coupled cavity. Assuming that the driver antenna is sensitive to the spatial phase of the fields \cite{Fink}, coupling to each mode will then have the spatial dependence $\cos{(k_{e}x_{d})}$ and $\cos{(k_{g}x_{d})}$, where $k_{e,g} = \omega_{e,g}/c$ and $x_{d}$ is the position of the driver antenna. The coupled wave system can be modeled by Maxwell equations,
\begin{widetext}
	\begin{equation}\label{cavity_eq1}
		\left\{
		\begin{array}{l}
		\displaystyle \left( \frac{\partial^2}{\partial t^2} - {c^2} \frac{\partial^2}{\partial x^2} \right) E_{e}(x_{d},t) = \chi\cos{(k_{e}x_{d})}\left( E_{e}(x_{r},t) + E_{g}(x_{r},t) \right)E_{p}(t)		\\ 	\\
		\displaystyle \left( \frac{\partial^2}{\partial t^2} - {c^2} \frac{\partial^2}{\partial x^2} \right) E_{g}(x_{d},t) = \chi\cos{(k_{g}x_{d})}\left( E_{e}(x_{r},t) + E_{g}(x_{r},t) \right)E_{p}(t)
		\end{array}	
		\right.	.
	\end{equation} 
\end{widetext}
Substituting in \Eq{cavity_eq1} the cavity modes $E_{e,g}(x_{d},t) = \cos{(k_{e,g}x_{d})}\mathcal{E}_{e,g}(t)$, the equation can be simplified with the rotating-wave approximation, 
\begin{equation}\label{cavity_eq2}
	\left\{
	\begin{array}{l}
	\displaystyle \frac{d^2}{d t^2} \mathcal{E}_{e}(t) = -\we^2 \mathcal{E}_{e}(t) + \chig\mathcal{E}_{g}(t)E_{p}(t)		\\ 	\\
	\displaystyle \frac{d^2}{d t^2} \mathcal{E}_{g}(t) = -\wg^2 \mathcal{E}_{g}(t) + \chie\mathcal{E}_{e}(t)E_{p}(t)
	\end{array}
	\right.	,
\end{equation}
where the effective nonlinear coupling parameters turn out to be $\chi_{e,g}(x_{r}) \equiv \chi\cos{(k_{e,g}x_{r})}$. Solutions to \Eq{cavity_eq2} will mimic \Eqs{OPA_sol}{DPA_sol} because \Eq{cavity_eq2} has the same form as \Eq{main_eq1}. Assuming resonant pumping, the amplification solutions are 
\begin{equation}
	\begin{split}
	E_{e}(t) &= E_{e}(0)\cos{(\we t)}\cosh{(\Omega_{0} t/2)},		\\
	E_{g}(t) &= \frac{\chi \cos{(k_{e} x_{r})}A_{0}E_{e}(0)}{2\wg\Omega_{0}}\sin{(\wg t)}\sinh{(\Omega_{0}t/2)}
	\end{split}
\end{equation}
where $\Omega_{0} = \sqrt{ \pm\chi^2\cos{(k_{e}x_{r})}\cos{(k_{g}x_{r})}A^2_{0}/4\we\wg }$, and initial conditions are $\tilde{E}_{e}(0) = E_{e}(0)$ and $\tilde{E}_{g}(0) = 0$. The $\pm$ sign in $\Omega_{0}$ corresponds to the sum-frequency pump $\nu = \Sw$ $(+)$ and the difference-frequency pump $\nu = \Dw$ $(-)$. As the wavelengths of the two cavity modes are slightly off, the nonlinear coupling can be either symmetric (same sign) or asymmetric (opposite sign) depending on the position of the receiver antenna $x_{r}$. For the symmetric case we say that the symmetry relation is positive ($\chie\chig > 0$), and for the asymmetric case the symmetry relation is negative ($\chie\chig < 0$). In the example of \Fig{fig:cavity}(c), the cavity is pumped with the difference frequency $\nu = \Dw$. As the receiver antenna moves across the spatial phase profiles of the cavity modes, amplification (DPA) occurs in the regimes 
\begin{equation}\label{DPA_regime}
	x_{r} \in \left( (2m+1)\lambda_{e}/4, (2m+1)\lambda_{g}/4 \right)	,
\end{equation}
where $\lambda_{e,g} = 2\pi/k_{e,g}$ and $m$ is an integer. In these regimes, the symmetry relation is negative $\chie\chig = \chi^2\cos{(k_{e}x_{r})}\cos{(k_{g}x_{r})}< 0$. Outside of these regimes, OPA can occur with a sum-frequency pump (See \Fig{fig:cavity}(b)). Dualism between DPA and OPA is manifested as the position of the receiver antenna changes the symmetry nature in the coupling. 

Finally, we address the problem of photon conservation in DPA. The standard Hamiltonian for second-order nonlinear interaction is $\hat{H}_{s} =  \hat{H}_{0} + \hbar \chi^{(2)} (\aue\aug \tilde{E}_{p} + \ade\adg \tilde{E}^{\ast}_{p} + \aue\adg \tilde{E}_{p} + \ade\aug \tilde{E}^{\ast}_{p})$ \cite{Gerry}, where $\chi^{(2)}$ is a real-valued parameter. The two terms $\aue\aug \tilde{E}_{p}$ and $\ade\adg \tilde{E}^{\ast}_{p}$ describe the anti-Jaynes-Cummings interaction that supports OPA. The photon picture for OPA is that one sum-frequency photon breaks into two lower energy photons at the target frequencies. The other two terms $ \aue\adg \tilde{E}_{p}$ and $\ade\aug \tilde{E}^{\ast}_{p}$ are the Jaynes-Cummings interaction which promotes energy exchange between the target fields. If we generalize the three-body Hamiltonian with nonlinear coupling parameters $\chie$ and $\chig$,
\begin{equation}\label{DPA_H}
	\begin{split}
	\hat{H}_{g} = \hat{H}_{0} + \hbar & (\chig\aue\aug \tilde{E}_{p} + \chie\ade\adg \tilde{E}^{\ast}_{p} 	\\
			   &+ \chig\aue\adg \tilde{E}_{p} + \chie\ade\aug \tilde{E}^{\ast}_{p}), 	
	\end{split}
\end{equation}
the generalized Hamiltonian $\hat{H}_{g}$ will yield a set of quantum Heisenberg equations that resemble \Eq{main_eq2}. Note that the standard Hamiltonian $\hat{H}_{s}$ is resumed by choosing $\chie = \chig = \chi^{(2)}$ in $\hat{H}_{g}$. The generalized Hamiltonian $\hat{H}_{g}$ in \Eq{DPA_H} leads to solutions for $\ade(t)$ and $\adg(t)$ similar to \Eqs{OPA_sol}{DPA_sol}. In particular, when the symmetry relation is negative $\chie\chig < 0$, a difference-frequency pump can give rise to an amplification solution. However, while the quantum solutions are similar to those from the classical analysis, expectation values of operators do not agree with corresponding classical observables. In the case of $\chie \ne \chig$, or $\chie = \chig = i\chi^{(2)}$, the generalized Hamiltonian $\hat{H}_{g}$ is non-Hermitian and the expectation value of total energy $\langle{H}_{g}\rangle$ is complex-valued. As a result, the photon picture is incompatible with DPA.

In conclusion, we derive from Maxwell equations the classical solutions for DPA as an alternative pathway of parametric amplification. In contrast to OPA, amplification in DPA requires a difference-frequency pump and negative symmetry relation of parametric coupling. We illustrate the dualism between OPA and DPA by showing their corresponding roles in mutually exclusive parameter regimes. We further show that generalizing the conventional OPA Hamiltonian to include DPA will lead to non-Hermiticity. This suggests a four-body Hamiltonian may be needed to construct a quantum mechanical system that properly describes DPA. As the DPA gain rate $\Wm = \sqrt{-\Dv^2-\chie\chig|\Ep|^2/4\we\wg}$ scales weakly with increasing target frequencies $\omega_{e,g}$, DPA could be suitable for efficient X-ray amplification.

\section{Acknowledgement}

The authors thank Da-Wei Wang, Luojia Wang, Xiwen Zhang, Anatoly Svidzinsky, Wolfgang Schleich and Marlan Scully for advice. W. C. Huang wishes to give special thanks to Steve Payne and William Seward for helpful discussions. This work utilized high-performance computing resources from the Holland Computing Center of the University of Nebraska. Funding for this work comes from NSF EPS-1430519 and NSF PHY-1602755.

\end{document}